# Origami-controlled strain engineering of tunable flat bands and correlated states in folded graphene

Li-Zhen Yang[1,3][†], Ling-Hui Tong[1,3][†], Cheng-Sheng Liao[1], Qilong Wu[1], Xiaoshuai Fu[1], Yue-Ying Zhou[1,3], Yuan Tian[1], Li Zhang[1], Lijie Zhang[1], Meng-Qiu Cai[1], Lin He[2], Zhihui Qin[1]*, and Long-Jing Yin[1,3]*

[1] *Key Laboratory for Micro/Nano Optoelectronic Devices of Ministry of Education & Hunan Provincial Key Laboratory of Low-Dimensional Structural Physics and Devices, School of Physics and Electronics, Hunan University, Changsha 410082, China*

[2] *Center for Advanced Quantum Studies, Department of Physics, Beijing Normal University, Beijing 100875, China*

[3] *Research Institute of Hunan University in Chongqing, Chongqing 401120, China*

[†] These authors contributed equally to this work

* Corresponding author: zhqin@hnu.edu.cn; yinlj@hnu.edu.cn

**Flat electronic bands with tunable structures offer opportunities for the exploitation and manipulation of exotic interacting quantum states. Here, we present a controllable route to construct easily tunable flat bands in folded graphene, by nano origami-controlled strain engineering, and discover correlated states in this system. Via tearing and folding graphene monolayer at arbitrary step edges with scanning tunneling microscope manipulation, we create strain-induced pseudo-magnetic fields as well as resulting flat electronic bands in the curved edges of folded graphene. We show that the intensity of the pseudo-magnetic field can be readily tuned by changing the width of the folding edge due to the edge-width-dependent lattice deformation, leading to the well adjustability of the geometry of flat bands in folded graphene. Furthermore, by creating expected dispersionless flat bands using this technique, the correlation-induced splits of flat bands are successfully observed in the density of states when these bands are partially filled. Our experiment provides a feasible and effective pathway to engineer the system with tunable flat band structures, and establishes a new platform that can be used to realize devisable strain and interaction induced quantum phases.**

Graphene-based flat band systems have recently attracted considerable attention owing to the emergence of rich correlation-driven electronic phases. The most two prominent examples are twisted graphene moiré superlattices and trilayer graphene/hexagonal boron nitride (hBN) heterostructures. In these two materials, the flat band induced correlated insulating, superconducting, ferromagnetic and topological states have been experimentally observed through carefully controlling their lattice structures [1-10]. For the graphene moiré superlattices, the interlayer twisted angle should approach a specific value of about 1.1°, *i.e.*, the so-called magic-angle, to efficiently narrow the bandwidth [1,2,11]. For the trilayer graphene/hBN system, the graphene layers should be *ABC*-stacked and precisely aligned to the hBN layers, so that the isolated flat minibands can be created [8-10]. These experimental fabrications remain challenging, leading to the difficulties in controlling the sample qualities and obtained results and also in tuning the electronic structures [3-5,11-13]. Another alternative way to construct flat bands is forming Landau level-like discrete bands by strain-induced pseudo-magnetic field (PMF) in graphene [14,15]. Such PMF, unlike the real magnetic field, does not break time-reversal symmetry and is expected to introduce superconductivity and other correlated states [16-19]. Although strain-induced PMF has been observed in several graphene structures [20-32], the controllable generation of scalable PMF and corresponding flat electronic bands with well adjustable configuration that can introduce correlated effects is still challenging.

In this Letter, we report an efficient route to create readily tunable PMF as well as induced flat bands and demonstrate the emergence of correlated states in folded graphene via nano origami-controlled strain engineering. Through the tip-manipulated tearing and folding of scanning tunneling microscope (STM), we controllably generate folded graphene structures at arbitrary monolayer step edges. The curved 1D edges of the folded graphene undergo stretched strains perpendicular to the edge, resulting in the formation of PMF confined in the 1D folding edge and the observation of a series of flat bands in the scanning tunneling spectroscopy (STS). We find that the intensity of the PMF can be easily modified by changing the width of the folding edge whose value is experimentally controlled by the folding area per unit length. This brings the well tunability of flat bands for the folded graphene system and leads to the observation of correlated states, opening a simple and feasible way to explore the flat band physics and strain-induced properties.

The folded graphene structures were constructed on graphite surface by using the STM tip-manipulated tearing and folding technique as developed in our recent work [33]. Briefly, the tip is significantly approached to graphene surface at the position below a step edge and then is straightly moved across the step edge along a predefined route. During this procedure, the greatly enhanced tip-graphene repulsive force can tear and fold graphene sheet partly at the step edge [34,35], leading to the formation of folded graphene nanostructures as schematically illustrated in Fig. 1(a). Based on this technique, we can controllably generate folded graphene not only from the specific nano-size graphene [Fig. 1(b)], but from arbitrary graphene sheets at step edges (Figs. 1(c) and 1(d) and see more experimental data in Fig. S1 [36])[37-51]. This allows us to create folded graphene structures without any other pretreatment of the graphene sample [52,53].

The obtained folded graphene structures provide a unique platform to explore strain-induced physics. With a close examination of the origami-created folded graphene, we can find that a bright and straight folding edge exists in the topographic images of the folded graphene samples (see Figs. 1(b)-1(d) and S1 [36]). This bright 1D folding edge is a curved structure and results from the lattice connection between the folding layer and the underlying graphene plane [52,54] (see schematic illustration in Fig. 1(a) and more experimental data in Fig. S2 [36]). Such a curved graphene folding structure is an unconventional configuration and has been expected to develop some interesting quantum phenomena including strain-induced electronic properties [55-57]. In our experiment, we find that nearly all of the folded graphene flakes created by the tip-manipulated tearing and folding have such 1D folding edge (see Fig. S1 for more experimental data [36]), indicating that the nano-origami technique is an efficient way to construct folded graphene with closed and curved folding edge associated with the induced novel quantum states [55-57]. The present work will show that such folding edge can introduce strain-governed PMF and flat electronic bands with tunable geometry.

The structure of the 1D folding edge can be well described by the folding angle and width. Here, the folding angle is defined as the angle between the armchair orientation of graphene lattices and the direction parallel to the folding edge [Fig. 2(a)]. Thus, the folding angles 0° and 30° correspond to the armchair and zigzag folding edges, respectively. Figure 2(b) shows the distribution of the folding angles obtained from 43 folded graphene. It clearly demonstrates that the graphene sheets are mainly folded along armchair or zigzag directions [about 42% (18) are armchair and 28% (12) are zigzag edges]. This preferred folding phenomenon is consistent well with that observed in ultrasound-induced folding of suspended graphene [54], which boils down to the minimal formation energy for the armchair and zigzag edges. The folding edge width ($w$) is obtained by measuring the full width at half maximum of the curved folding edge from the topographic image [see inset in Fig. 2(a)]. The measured width of different folding edges approximately ranges from 2 to 6 nm as shown in Fig. 2(c).

The most striking feature observed in the origami-created folded graphene is the existence of non-uniform strain in the 1D folding edge region. This can be obtained by analyzing local lattice deformation from the STM topographic images and the corresponding fast Fourier transforms in different positions of the folded graphene sheet [27-30,58]. We find that there are stretched strains perpendicular to the folding edge direction in the curved 1D folding edges. That is to say, for the 0° and 30° folding edges, the graphene lattices are stretched along zigzag and armchair directions respectively, as exemplified in Figs. 2(d), 2(e) and S3 [36]. In order to quantitate the strain observed in the curved 1D edge, we measured the relative lattice deformation by comparing the graphene lattice perpendicular to the folding edge to that along the folding edge (see Figs. S3 and S4 for more discussion [36]). The results are shown in Fig. 2(f) as a plot of relative lattice deformation versus $w$. Non-uniform strains exist obviously in the measured folded graphene. Interestingly, it is seen that the obtained relative lattice deformation decreases with increasing $w$. This means that the intensity of non-uniform strain in the folded graphene can be modified by varying the width of the folding edge. We suspect that such a width dependence of strain is attributed to the strain relaxation in folded graphene, which is discussed detailedly in Supplemental Material [36].

Theoretically, a non-uniform strain will introduce an effective gauge potential through altering the electron hopping energies in graphene, thereby resulting in the creation of PMF whose intensity is proportional to the strain field [14,15]. This can be verified by local spectroscopic measurements with spatial resolution in the strained graphene. The strain-induced PMF can effectively modulate the low-energy band structures by creating quantized pseudo-Landau levels (PLLs) which can be directly observed in the STS spectrum [*i.e.*, the differential conductivity (*dI/dV*) spectrum]. The energies $E_n$ of the PLLs with different indices $n$ follow the sequence of massless Dirac fermion LLs in monolayer graphene [59-61] as described by $E_n = E_D + \mathrm{sgn}(n)\sqrt{2e\hbar v_F^2 |n| B_S}$ (here $E_D$ is the energy of Dirac point, $e$ is the electron charge, $\hbar$ is the reduced Planck's constant, $v_F$ is the Fermi velocity, and $B_S$ is the magnitude of PMF). Figure 3(b) shows the spatially resolved STS spectra measured in a folded graphene with $w$ = 5.6 nm [Fig. 3(a)]. The tunneling spectra recorded on the 1D folding edge exhibit a series of conductance peaks with unequal energy separations which arise from PMF-induced PLLs (other possible origins such as the 1D electron confinement [62] can be excluded, see Supplemental Material [36] for details). The peak features are absent in the STS spectra taken at the flat region of folded graphene and graphite substrate, which both show a typical V-shaped structure of graphene. This phenomenon is observed in many other folded graphene samples with varying $w$. Figure 3(c) shows representative tunneling spectra measured on four folding edges with different widths (4.0 nm, 5.0 nm, 5.6 nm and 7.5 nm). The PLL peaks are clearly displayed in all tunneling spectra of Fig. 3(c). Through labeling the corresponding PLL indices $n$ (starting from the $n$ = 0 level at the charge neutrality point), the extracted peak energies exhibit a well linear dependence on $\mathrm{sgn}(n)(|n|)^{1/2}$ [Fig. 3(d)]. These results unambiguously confirm the existence of strain-induced PMF as well as resulting PLLs in the curved folding edges.

The magnitude of PMF, $B_S$, can be directly estimated by fitting the PLL peak sequence to the above-mentioned theoretical equation. Figure 3(e) summarizes the obtained $B_S$ for the folded graphene with varying $w$ (see Fig. S5 for more discussion [36]). For a certain folded graphene, it exhibits a relatively uniform PMF in the folding edge (see Figs. S6 and S7 [36]). For different folded graphene, $B_S$ ranges from 10 T to nearly 60 T. The obtained value of PMF is consistent with that reported in graphene ripples with similar geometries both theoretically [63,64] and experimentally [24,26]. More interestingly, $B_S$ exhibits a monotonically decreasing feature as increasing $w$. In other words, the intensity of PMF has the same $w$-dependence as the lattice deformation shown in Fig. 2(f). This result is quite reasonable because the strain-induced PMF is proportional to the strain level [14,15], further supporting the interpretation of strain-induced PMF in the folded graphene. The above results also indicate that the intensity of PMF as well as the structure of the resulting PLLs (including the energy location and separation and even the bandwidth [19]) can be well tuned by changing $w$ in the origami-created folded graphene. Previously, strain-induced PMF has been experimentally generated in some special graphene structures, such as nanobubbles [20-22], ripples [23-26], heterostructures [19,27,28], and nanopillar-supported graphene [29,30]. However, a controllable way to efficiently construct PMF with tunable intensity is still very lacking. The origami-controlled graphene tearing and folding presented in this work thus provides such a desired route to create tunable PMF, enabling the easy generation of flat discrete bands with prescribed configuration and thus the exploration of correlation-induced electronic states.

The tunability of the PMF in the origami-created folded graphene can be further revealed by investigating the experimental controllability of $w$. In the experiment, we find that $w$ shows a roughly inversely proportional dependence on the folding area per unit folding-edge-length (Figs. S8 and S9 [36]). This means that $w$ can be controlled by altering the size of the folded graphene sheet per unit length. Experimentally, operating different times tip-manipulated origami processes along the same tip moving path can change the folding area per unit length (as shown in Fig. S10 [36] and as reported in our previous work [33]), which in turn controls $w$ and the emergent PMF. The origin of the dependence between $w$ and folding area can be understood by the calculation of graphene folding energetics (see Supplemental Material [36] for details). The above results demonstrate explicitly that the origami-created folded graphene system has a well feasibility and controllability for constructing strain-induced PMF with tunable intensity.

Based on the above results, we now explore correlation effects of flat bands in folded graphene by creating appropriate strain-induced dispersionless flat bands using the tip-induced origami technique under liquid helium temperature. Figure 4(a) shows an origami-created folded graphene structure with $w = 4.5$ nm obtained at 4.6 K (see Fig. S11 for more details [36]). Similarly, the STS spectrum measured on the folding edge region [Figs. 4(b) and 4(c)] exhibits a sequence of PMF-induced PLL peaks. The value of the PMF is extracted ~13 T by fitting [inset of Fig. 4(b)]. Besides, there are three notable features observed in the STS spectra of the folding edge. First, a pronounced $n = 0$ level peak exists near the Fermi energy with a smallest bandwidth of ~14 meV. This width is slightly narrower than that of the flat bands (~18 meV) detected in the magic-angle twisted bilayer graphene under the same experimental conditions [65], suggesting that the kinetic energy of quasiparticles in this flat band is smaller enough for the generation of strongly correlated states. Second, negative differential conductance is observed between $n = 0$ and $n = 1$ levels [Figs. 4(b) and 4(c)], evidencing the well separation between $n = 0$ and higher levels (see Fig. S12 for more discussion [36]). Third, there is a slight change of doping in different edge positions, which is reflected by the variation of filling state of the $n = 0$ PLL flat band: from full filling to partial filling [Fig. 4(c) and see Supplemental Material [36] for more discussion]. Thanks to the above three features, we observed clear correlated effects in this sample. As shown in Fig. 4(c), when the $n = 0$ PLL is fully filled, it shows a single sharp peak in the tunnelling spectrum (top curve). While when the $n = 0$ level is partially filled, the peak splits into two shoulder peaks with varying relative intensities for different partial filling states (lower three curves). Particularly, the two splitting peaks exhibit equal intensities at nearly half-filling state accompanying an energy separation of ~15 meV (red curve). The cooccurrence of the partial-filling-induced spectroscopic splitting and peak weight redistribution in the PLL flat bands is a clear manifestation of the emergence of correlated states in the folded graphene, which is very analogous to that obtained in the flat bands of magic-angle twisted bilayer graphene displaying correlated insulating and superconducting phases [66-68]. A similar phenomenon has also been observed in strain-induced flat bands of graphene superlattices [19,32], where a valley-polarized ground state with a comparable gap was detected [32]. Further evidence about the existence of correlated effects in the folded graphene is obtained by examining the quasiparticle lifetime, from which a interaction-induced linear energy dependence of inverse lifetime is observed [69-71] (see Supplemental Material [36] for details). Our results demonstrate that the origami-created folded graphene is indeed a feasible flat-band system that can host correlated electronic states, and more importantly, it shows a well tunability of the flat bands which can be used to manipulate the correlation effects.

Although the geometry is different between our sample and the graphene moiré systems, they exhibit very similar interaction-induced spectroscopic characteristics as discussed above. A similar situation has also been recently observed in ABC-stacked trilayer graphene system in which both ferromagnetism and superconductivity have emerged with [8,10] and without [72,73] the existence of a moiré superlattice. It seems that the ultra-narrow bands are the most important role. The observed correlated effects in our folded graphene thus further support such a view, providing another non-superlattice and structure simpler strongly correlated graphene system. In addition, considering the specific structure of our folded graphene, the underlying microscopic mechanism of the observed correlated effects may also be related to the quasi-1D feature of the folding edge. Further experiments are needed to understand the origin of the correlated states as well as reveal more exotic quantum states in folded graphene.

In conclusion, we have demonstrated a controllable method to create easily tunable flat bands and generated correlated states in folded graphene via nano origami-controlled strain engineering. Our work indicates that the origami-created folded graphene system is an attractive platform that can realize devisable strain and band flattening engineering, offering a simple and effective approach to construct strain- and correlation-driven quantum phases with high tunability. The investigation of this non-moiré flat-band system may help us to better understand the correlated-electron physics in graphene. Our experiment is also instrumental in the exploitation of strain-modulated emergent properties in other van der Waals layered materials.

## Acknowledgements

The authors thank H. Jiang for helpful discussion. This work was supported by the National Natural Science Foundation of China (Grant Nos. 12174095, 11804089, 12174096, 51772087, 11904094 and 51972106), the Natural Science Foundation of Hunan Province, China (Grant No. 2021JJ20026), and the Strategic Priority Research Program of Chinese Academy of Sciences (Grant No. XDB30000000). L.J.Y. also acknowledges support from the Science and Technology Innovation Program of Hunan Province (Grant No. 2021RC3037) and the Natural Science Foundation of Chongqing, China (cstc2021jcyj-msxmX0381). The authors acknowledge the financial support from the Fundamental Research Funds for the Central Universities of China.

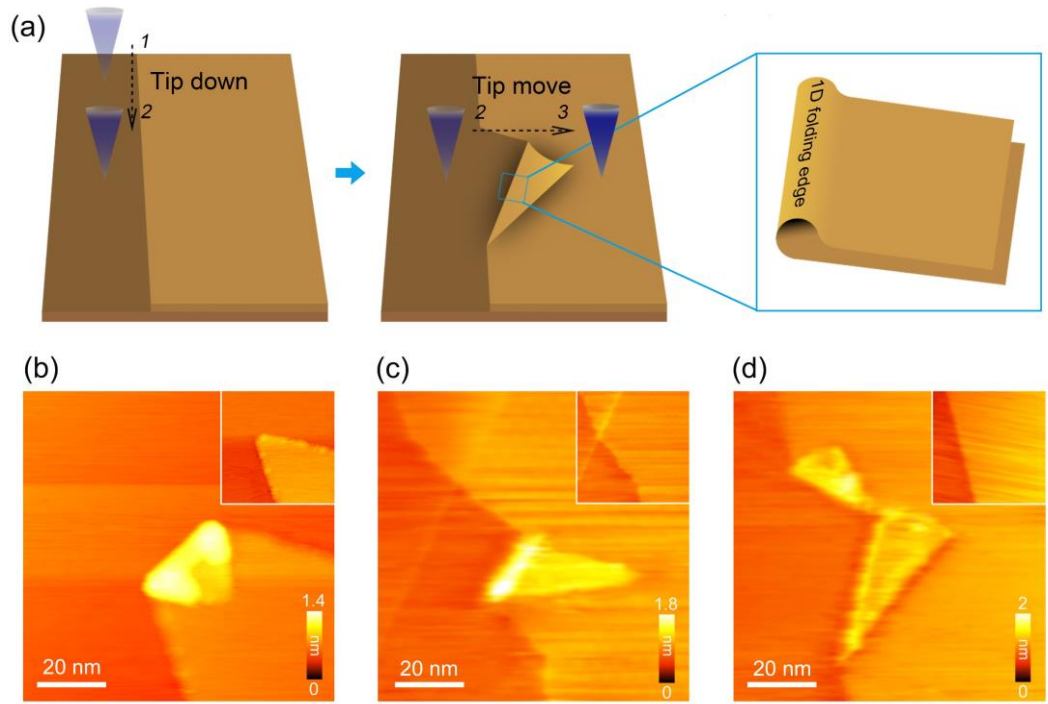

FIG. 1. (a) Schematic illustration of the tip-induced graphene tearing and folding process at a step edge. The blue frame shows the zoom-in structure of the folding edge. (b)-(d) Representative STM topographic images of three folded monolayer graphene structures created by tip-induced origami on graphite surface. Insets are original STM topographic images for each folded graphene. Tunnelling parameters: $V_b$ = 0.5 V, $I$ = 0.2 nA (b); $V_b$ = 0.6 V, $I$ = 0.2 nA (c); $V_b$ = -0.3 V, $I$ = 0.2 nA (d).

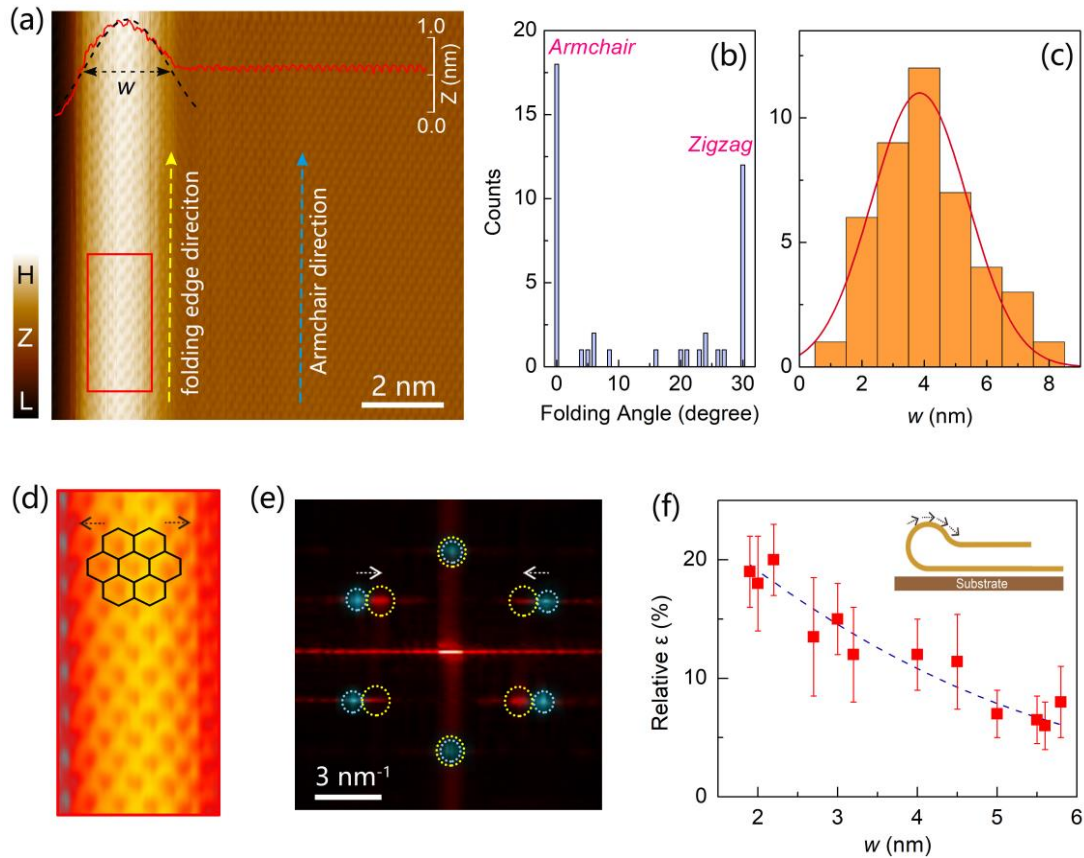

FIG. 2. (a) Atomic-resolution STM image ($V_b$ = -0.2 V, $I$ = 0.2 nA) around the folding edge of a folded graphene. The bright region represents the 1D folding edge. Arrows indicate directions along the 1D folding edge and armchair orientation of graphene. Inset: height profile across the folding edge. The folding edge width ($w$) is extracted by the Gaussian fitting (dashed curve) to the height profile of the folding edge region. (b) Distribution of the folding angle obtained from 43 folded graphene. (c) Distribution of $w$. The red curve is a Gaussian fit to the data. (d) Zoom-in atomic-resolution image of the red frame in (a). The graphene hexagonal lattice is superimposed with part of the image. The lattice is clearly stretched along the zigzag direction. (e) Superimposed Fourier transforms of the atomic structures in the folding edge region (yellow dashed circle) and the flat region (blue dashed circle) in (a). The lattice of the folding edge is stretched perpendicular to the edge direction compared to that of the flat region. (f) Relative lattice deformation ($\varepsilon$) as a function of $w$. The dashed line is a guide to show observed trend. Inset is schematic side view of folded graphene edge.

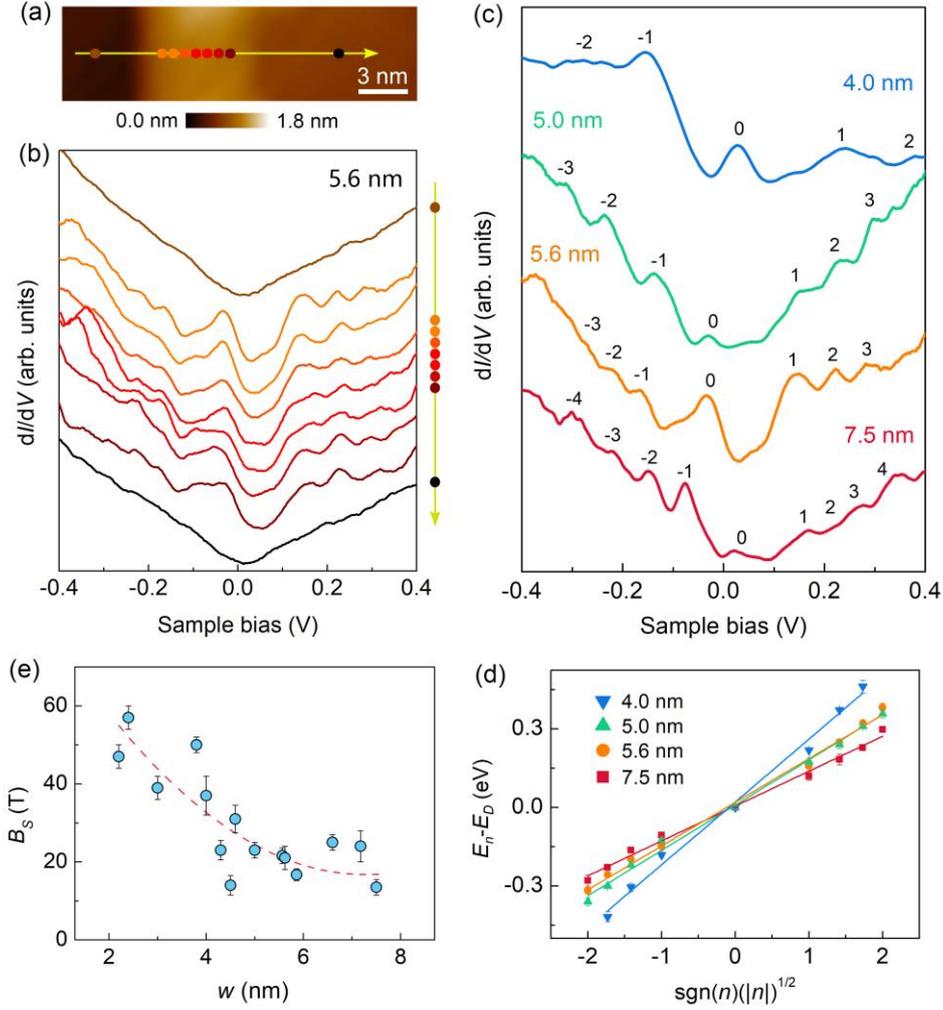

FIG. 3. (a) STM topographic image of a folded graphene with $w$ = 5.6 nm. (b) Spatially resolved $dI/dV$ spectra obtained at the positions marked by the dots along the arrow in (a). (c) Typical $dI/dV$ spectra measured on the top of the 1D folding edges with $w$ = 4.0 nm, 5.0 nm, 5.6 nm, and 7.5 nm. The spectra show quantized conductance peaks (labeled by corresponding PLL index) and are shifted vertically for clarity. The extra peak-like feature near zero bias, such as the one around 55 mV for the 5.6 nm sample, may arise from the finite coupling to the substrate [69] or the inelastic tunneling signature that mainly exists in quasi-freestanding graphene [74-76] (see Supplemental Material [36] for more discussion). (d) PLL peak energies (dots) extracted from (c) as a function of $\text{sgn}(n)(|n|)^{1/2}$. Lines are linear fits to the data. (e) Measured $B_S$ plotted versus $w$. The dashed line is a guide to the eye.

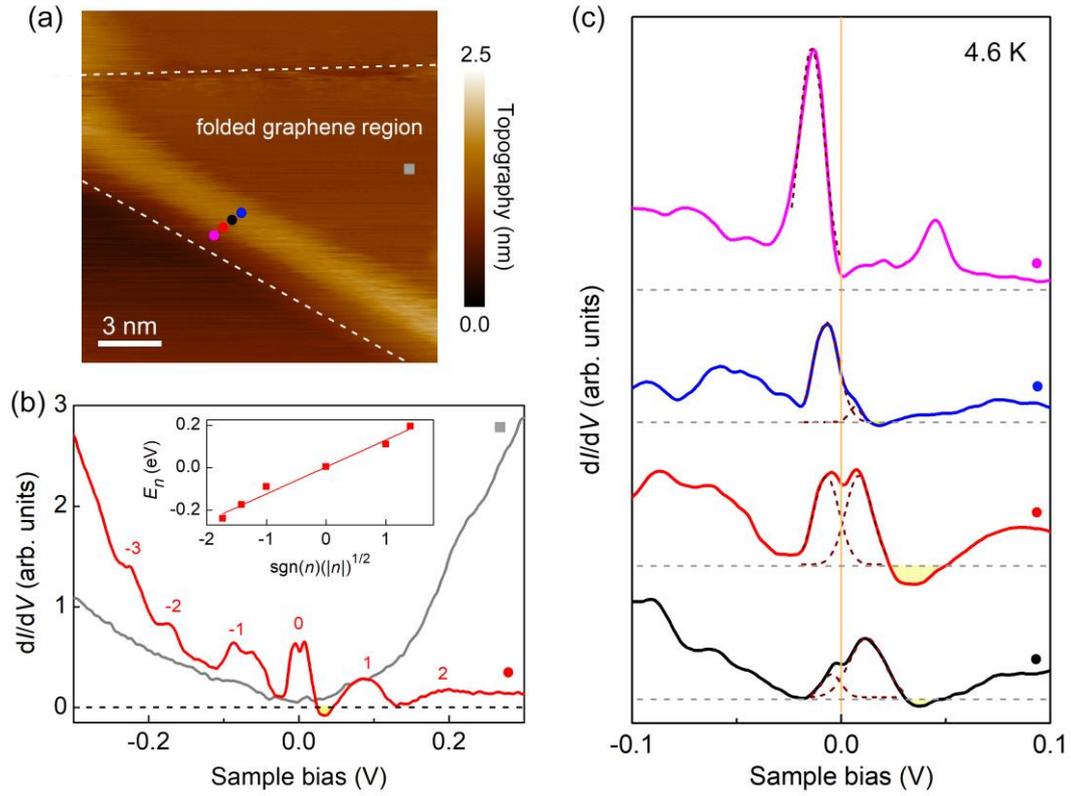

FIG. 4. (a) STM image ($V_b$ = 0.5 V, $I$ = 0.2 nA) of a folded graphene on graphite. (b) Typical STS spectra for the folding edge region (red curve) and flat region (gray curve) taken at the positions marked by the red dot and gray square in (a), respectively. Inset: PLL energies of the folding edge region plotted versus sgn($n$)($|n|$)$^{1/2}$. (c) STS spectra recorded at different positions of the folding edge labelled by the colored dots in (a) showing different fillings of the flat band near Fermi energy. The dashed peak curves are peak fits to the flat bands. The dashed straight lines denote the position of zero conductance for each spectrum.